\documentclass{PoS}
\usepackage{bm}
\usepackage{wrapfig}
\newcommand{\be}{\begin{equation}}
\newcommand{\ee}{\end{equation}}
\newcommand{\bea}{\begin{eqnarray}}
\newcommand{\eea}{\end{eqnarray}}

\newcommand{\bfk}{\bm{k}}
\def\bkt{\bfk_\perp}

\newcommand{\bfp}{\bm{p}}
\def\bpp{\bfp_\perp}

\newcommand{\bfP}{\bm{P}}
\newcommand{\bfS}{\bm{S}}
\newcommand{\bfs}{\bm{s}}

\newcommand{\pup}{p^\uparrow}

\newcommand{\qup}{q^\uparrow}

\def\lsim{\mathrel{\rlap{\lower4pt\hbox{\hskip1pt$\sim$}}\raise1pt\hbox{$<$}}}

\title{$A_N$ in inclusive lepton-proton collisions}

\ShortTitle{$A_N$ in inclusive lepton-proton collisions}

\author{M.~Anselmino, M.~Boglione\\
Dipartimento di Fisica Teorica, Universit\`a di Torino, and INFN, Sezione di Torino, \\
Via P.~Giuria~1, I-10125 Torino, Italy\\
E-mail: \email{anselmino@to.infn.it}, \email{boglione@to.infn.it}}

\author{\speaker{U.~D'Alesio}\\
       Dipartimento di Fisica, Universit\`a di Cagliari, and INFN, Sezione di Cagliari,  C.P.~170, I-09042 Monserrato (CA), Italy \\
        E-mail: \email{umberto.dalesio@ca.infn.it}}

\author{S.~Melis\\
Dipartimento di Fisica Teorica, Universit\`a di Torino, Via P.~Giuria~1, I-10125 Torino, Italy\\
E-mail: \email{melis@to.infn.it}}

\author{F.~Murgia\\
       INFN, Sezione di Cagliari,  C.P.~170, I-09042 Monserrato (CA), Italy \\
        E-mail: \email{francesco.murgia@ca.infn.it}}

\author{A.~Prokudin\\
       Jefferson Laboratory, 12000 Jefferson Avenue, Newport News, VA 23606, USA\\
       E-mail: \email{prokudin@jlab.org}}

\abstract{Some estimates for the transverse single spin asymmetry, $A_N$, in the inclusive processes $\ell p^\uparrow \to h\,X$ are compared with new experimental data. The calculations are based on the Sivers and Collins functions as extracted from SIDIS azimuthal asymmetries, within a transverse momentum dependent factorization approach. The values of $A_N$ thus obtained agree in sign and shape with the data. Predictions for future experiments are also given.}

\FullConference{XXII. International Workshop on Deep-Inelastic Scattering and Related Subjects,\\
		28 April - 2 May 2014\\
		Warsaw, Poland}

\begin{document}

\section{Introduction and Formalism}

We present a phenomenological analysis of recent HERMES data~\cite{Airapetian:2013bim} for the single spin asymmetry (SSA) measured in the inclusive hadron production in lepton proton collisions. This study is based on a previous paper~\cite{Anselmino:2009pn}, recently extended~\cite{Anselmino:2014eza}, where we considered the transverse SSAs for the $\ell \, \pup \to h \, X$ process in the $\ell-p$ center of mass ({\it c.m.}) frame, with a single large $P_T$ final particle.

Such $A_N$ is the exact analogue of the SSAs observed in $p \, \pup \to h \, X$, the well known and large left-right asymmetries (see Ref.~\cite{D'Alesio:2007jt} and references therein). On the other hand, the process is essentially a semi-inclusive deep inelastic scattering (SIDIS) process, for which, at large $Q^2$ values (and small $P_T$ in the $\gamma^* - p\,$ {\it c.m.} frame), the TMD factorization is proven to hold~\cite{Collins:2011book,GarciaEchevarria:2011rb}. Notice that even without the detections of the final lepton, large $P_T$ values imply large values of $Q^2$.

We computed these SSAs assuming the TMD factorization and using the relevant TMDs (Sivers and Collins functions) as extracted from SIDIS data.
A first simplified study of $A_N$ in $\ell \, \pup \to h \, X$ processes was performed in Ref.~\cite{Anselmino:1999gd}. The process was also considered in Refs.~\cite{Koike:2002ti} in the framework of collinear twist-three formalism.

In Ref.~\cite{Anselmino:2009pn} (where all details can be found) we considered the process $\pup \ell \to h \, X$ in the proton-lepton {\it c.m.} frame (with the polarized proton moving along the positive $Z_{cm}$ axis) with:
\be
A_N = \frac{d\sigma^\uparrow(\bfP_T) - d\sigma^\downarrow(\bfP_T)}
           {d\sigma^\uparrow(\bfP_T) + d\sigma^\downarrow(\bfP_T)}
    = \frac{d\sigma^\uparrow(\bfP_T) - d\sigma^\uparrow(-\bfP_T)}
           {2 \, d\sigma^{\rm unp}(\bfP_T)} \,, \label{an}
\ee
where
\be
d\sigma^{\uparrow, \downarrow} \equiv \frac{E_h \, d\sigma^{p^{\uparrow,
\downarrow} \, \ell \to h\, X}}{d^{3} \bfP_h}
\ee
is the cross section for the inclusive process $p^{\uparrow, \downarrow}\, \ell \to h \, X$ with a transversely polarized proton with spin $\uparrow$ or $\downarrow$ w.r.t.~the scattering
plane~\cite{Anselmino:2009pn}.
For a generic transverse polarization along an azimuthal direction $\phi_S$ in the chosen reference frame, in which the $\uparrow$ direction is given by $\phi_S = \pi/2$, one has:
\be
A(\phi_S, S_T) = \bfS_T \cdot (\hat{\bfp} \times \hat{\bfP}_T) \, A_N =
S_T \sin\phi_S \, A_N \>, \label{phis}
\ee
where $\bfp$ is the proton momentum. Notice that one simply has:
\be
A_{TU}^{\sin\phi_S} \equiv \frac{2}{S_T} \,
\frac{\int \, d\phi_S \> [d\sigma(\phi_S) - d\sigma(\phi_S + \pi)]\> \sin\phi_S}
     {\int \, d\phi_S \> [d\sigma(\phi_S) + d\sigma(\phi_S + \pi)]}
= A_N \>.
\label{ATU}
\ee
Within a TMD factorization scheme for the process $p \,\ell \to h \, X$ with a single large scale (the final hadron transverse momentum $P_T$ in the proton-lepton {\it c.m.}~frame) the main contribution to $A_N$ comes from the Sivers and Collins effects~\cite{Anselmino:2009pn}:
\be
A_N =
\frac
{{\displaystyle \sum_{q,\{\lambda\}} \int \frac{dx \, dz}
{16\,\pi^2 x\,z^2 s}}\;
d^2 \bfk_{\perp} \, d^3 \bfp_{\perp}\,
\delta(\bfp_{\perp} \cdot \hat{\bfp}'_q) \, J(p_\perp)
\> \delta(\hat s + \hat t + \hat u)
\> [\Sigma(\uparrow) - \Sigma(\downarrow)]^{q \ell \to q \ell}}
{{\displaystyle \sum_{q,\{\lambda\}} \int \frac{dx \, dz}
{16\,\pi^2 x\,z^2 s}}\;
d^2 \bfk_{\perp} \, d^3 \bfp_{\perp}\,
\delta(\bfp_{\perp} \cdot \hat{\bfp}'_q) \, J(p_\perp)
\> \delta(\hat s + \hat t + \hat u)
\> [\Sigma(\uparrow) + \Sigma(\downarrow)]^{q \ell \to q \ell}} \>,
\label{anh}
\ee
with
\bea
\sum_{\{\lambda\}}\,[\Sigma(\uparrow) - \Sigma(\downarrow)]^{q
\ell \to q \ell} &=& \frac{1}{2} \, \Delta^N\! f_{q/\pup}
(x,k_{\perp}) \cos\phi \, \left[\,|{\hat M}_1^0|^2 + |{\hat
M}_2^0|^2 \right] \,
D_{h/q} (z, p_{\perp})  \nonumber \\
&+& h_{1q}(x,k_{\perp}) \, \hat M_1^0 \hat M_2^0 \, \Delta^N\!
D_{h/\qup} (z, p_{\perp}) \, \cos(\phi' + \phi_q^h) \label{ds1}
\eea
%
%
\be
\sum_{\{\lambda\}}\,[\Sigma(\uparrow) +
\Sigma(\downarrow)]^{q \ell \to q \ell} =
f_{q/p} (x,k_{\perp}) \,
\left[\,|{\hat M}_1^0|^2 + |{\hat M}_2^0|^2 \right] \,
D_{h/q} (z, p_{\perp}) \>. \label{ss1}
\ee
All details can be found in Refs.~\cite{Anselmino:2009pn, Anselmino:2014eza}. Here we simply recall some main features.
\begin{itemize}
\item
$\bkt, \bpp$ are respectively the transverse momenta of the parton in the proton and of the final hadron w.r.t.~the direction of the fragmenting parton, with momentum $\bfp^\prime_q$. $\phi$ is the azimuthal angle of $\bkt$.
\item
The first term on the r.h.s.~of Eq.~(\ref{ds1}) shows the contribution of the Sivers effect~\cite{Sivers:1989cc,Bacchetta:2004jz},
\be
\Delta \hat f_{q/p,S}(x, \bfk_{\perp}) \equiv \Delta^N\! f_{q/\pup}\,(x, k_{\perp}) \>
\bfS_T \cdot (\hat{\bfp} \times \hat{\bfk}_{\perp }) \label{defsivnoi}
= -2 \, \frac{k_\perp}{M} \, f_{1T}^{\perp q}(x, k_{\perp}) \>
\bfS_T \cdot (\hat{\bfp} \times \hat{\bfk}_{\perp }) \>.
\ee
It couples to the unpolarized elementary interaction ($\propto (|\hat M_1^0|^2 + |\hat M_2^0|^2)$) and the unpolarized fragmentation function $D_{h/q} (z, p_{\perp})$; the $\cos\phi$ factor arises from the $\bfS_T \cdot (\hat{\bfp} \times \hat{\bfk}_{\perp })$ factor.
\item
The second term on the r.h.s.~of Eq.~(\ref{ds1}) represents the contribution to $A_N$ of the unintegrated transversity distribution $h_{1q}(x,k_{\perp})$ coupled to the Collins function $\Delta^N\! D_{h/\qup} (z,p_{\perp})$~\cite{Collins:1992kk,Bacchetta:2004jz},
\bea
\Delta \hat D_{h/q^\uparrow}\,(z, \bfp_{\perp}) \equiv \Delta^N\! D_{h/\qup}\,(z, p_{\perp}) \>
\bfs_q \cdot (\hat{\bfp}_q^\prime \times \hat{\bfp}_{\perp })
\label{defcolnoi} = \frac{2 \, p_\perp}{z \, m_h} H_{1}^{\perp q}(z,
p_{\perp}) \> \bfs_q \cdot (\hat{\bfp}_q^\prime \times
\hat{\bfp}_{\perp}) \>.
\eea
This effect couples to the spin transfer elementary interaction ($d\hat\sigma^{q^\uparrow \ell \to q^\uparrow \ell} - d\hat\sigma^{q^\uparrow \ell \to q^\downarrow \ell} \propto \hat M_1^0 \, \hat M_2^0$). The factor $\cos(\phi' + \phi_q^h)$ arises from phases in the $\bfk_\perp$-dependent transversity distribution, the Collins function and the elementary polarized interaction.
\end{itemize}

In HERMES paper~\cite{Airapetian:2013bim} the lepton moves along the positive $Z_{cm}$ axis. In this reference frame the $\uparrow$ ($\downarrow$) direction is still along the $+Y_{cm}$ ($-Y_{cm}$) axis as in Ref.~\cite{Anselmino:2009pn} and only the sign of $x_F = 2 P_L/\sqrt s$ is reversed.
More precisely the HERMES azimuthal dependent cross section is defined as~\cite{Airapetian:2013bim}:
\be
d\sigma = d\sigma_{UU}[1+S_T \, A_{UT}^{\sin\psi} \sin\psi] \>,\>\>
\label{sigH}
%
{\rm where}\>\>\>\>
\sin \psi = \bfS_T \cdot (\hat{\bfP}_T \times \hat{\bfk})
\ee
coincides with our $\sin\phi_S$ of Eq.~(\ref{phis}), as $\bfp$ and $\bfk$ (the lepton momentum) are opposite vectors
in the lepton-proton {\it c.m.}~frame. Taking into account that ``left" and ``right" are interchanged in Refs.~\cite{Anselmino:2009pn} and \cite{Airapetian:2013bim} (being defined looking downstream along opposite directions, $\bfp$ and $\bm{k}$) and the definition of $x_F$, one has:
\be
A_{UT}^{\sin\psi}(x_F, P_T) = A_N^{\pup \ell \to h X}(-x_F, P_T)  \>,
\label{AUT-hermes}
\ee
where $A_N^{\pup \ell \to h X}$ is the SSA in Eq.~(\ref{anh})~\cite{Anselmino:2009pn}, and $A_{UT}^{\sin\psi}$ is the quantity measured by HERMES~\cite{Airapetian:2013bim}.

\section{Results}
In the following, adopting the HERMES notation, we show our estimates based on two representative extractions of the Sivers and Collins functions:
$i)$  the Sivers functions from Ref.~\cite{Anselmino:2005ea} (only up and down quarks), together with the first extraction of the transversity and Collins functions of Ref.~\cite{Anselmino:2007fs} (SIDIS~1 in the following). In such studies the Kretzer set for the collinear fragmentation functions (FFs)~\cite{Kretzer:2000yf} was adopted;
$ii)$ the Sivers functions from Ref.~\cite{Anselmino:2008sga}, where also the sea quark contributions
were included, together with an updated extraction of the transversity and Collins functions~\cite{Anselmino:2008jk} (SIDIS~2 in the following); in these cases we adopted another set for the FFs, namely that one by de Florian, Sassot and Stratmann (DSS)~\cite{deFlorian:2007aj}.

We consider both the fully inclusive measurements $\ell \, p \to \pi \, X$ at large $P_T$, as well as the sub-sample of data in which also the final lepton is tagged (SIDIS category).
In the first case the only large scale is the $P_T$ of the final pion, and for $P_T \simeq 1$ GeV, to avoid the low $Q^2$ region, one has to look at pion production in the backward proton hemisphere, ($x_F>0$ in the
HERMES conventions). For the tagged-lepton sub-sample data $Q^2$ is always bigger than 1 GeV$^2$ and $P_T$ is still defined w.r.t.~the lepton-proton direction. 

In both cases (inclusive or SIDIS events) the Sivers and Collins effects are not separable.

\noindent
$\bullet$ Fully inclusive case

\begin{figure}[h!t]
\includegraphics[width=3.9truecm,angle=0]{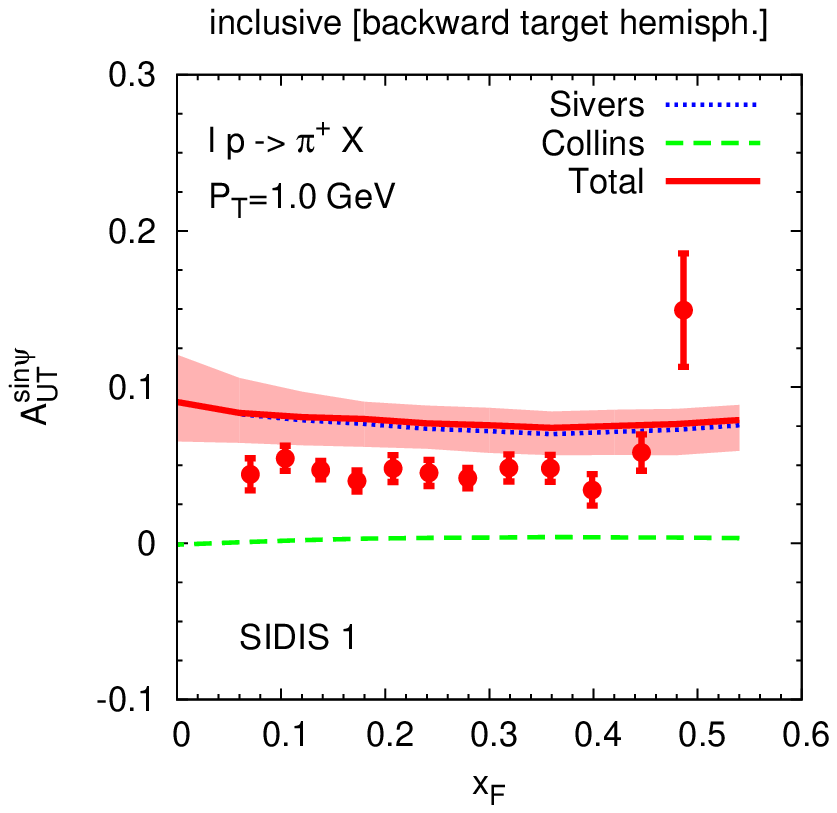}
\hspace*{-8pt}\includegraphics[width=3.9truecm,angle=0]{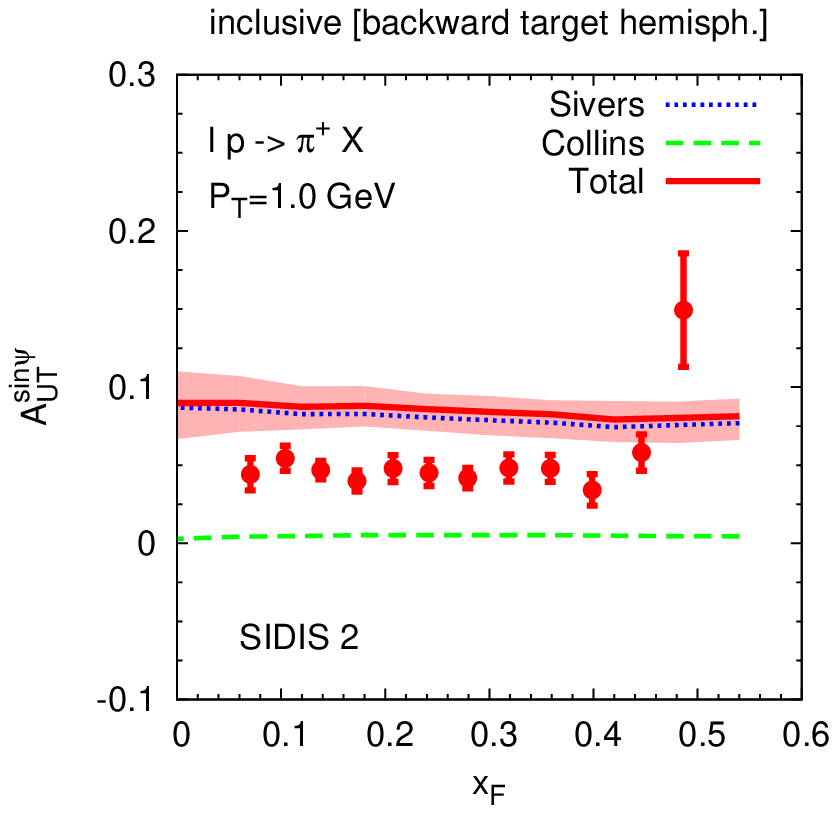}
\hspace*{-8pt}\includegraphics[width=3.9truecm,angle=0]{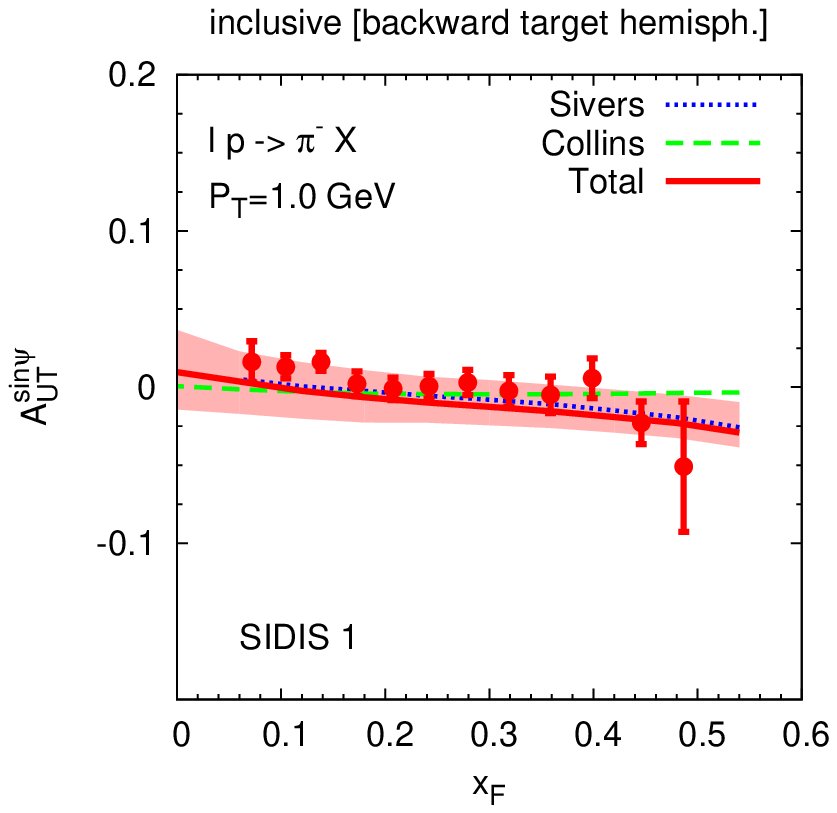}
\hspace*{-8pt}\includegraphics[width=3.9truecm,angle=0]{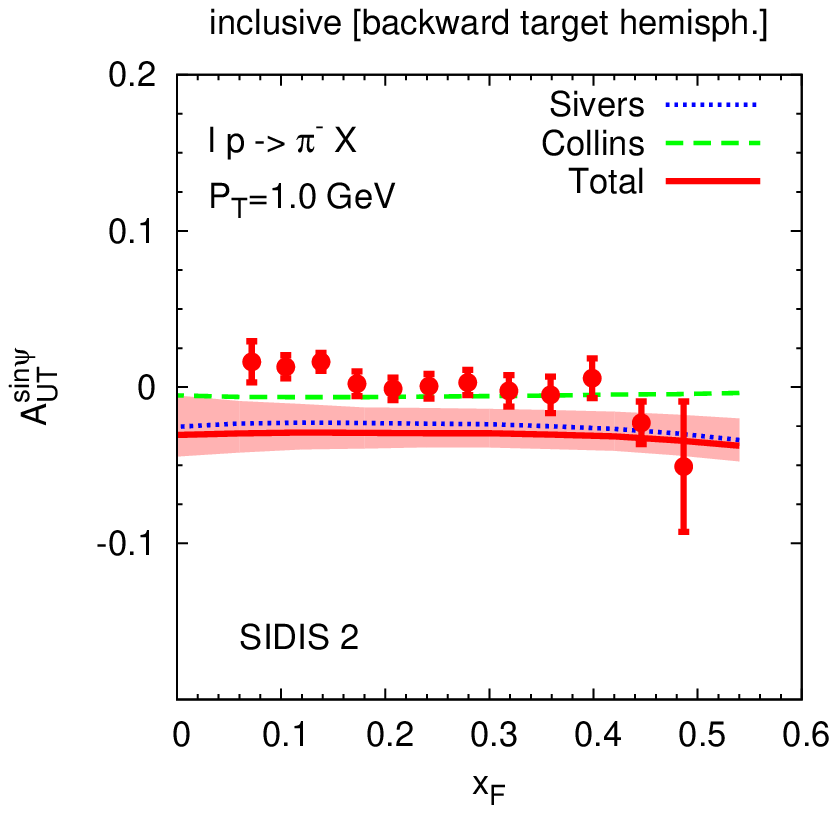}
\caption{Theoretical estimates for $A_{UT}^{\sin\psi}$ vs.~$x_F$ at $P_T = 1$ GeV for inclusive $\pi^+$ (first and second panel) and $\pi^-$ (third and fourth panel) production in $\ell \, \pup \to \pi \, X$ processes, computed according to Eqs.~(\protect\ref{AUT-hermes}) and (\protect\ref{anh})--(\protect\ref{ss1}) of the text and compared with the HERMES data~\cite{Airapetian:2013bim}. See the legend and text for details.
\label{fig:an-hermes-pi}}
\end{figure}

Only one HERMES data bin covers moderately large $P_T$ values, with $1 \lsim P_T \lsim 2.2$ GeV, and $\langle P_T \rangle \simeq$ 1-1.1 GeV. In Fig.~\ref{fig:an-hermes-pi} we show the results for $\pi^+$ (first and second panel) and $\pi^-$ (third and fourth panel) production coming from SIDIS 1 and SIDIS 2 sets, for the Sivers (dotted blue lines) and Collins (dashed green lines) effects separately, together with their sum (solid red lines) and the envelope of the statistical error bands (shaded area): $i)$ here the Collins effect is almost zero, as the partonic spin transfer in the backward proton hemisphere is dynamically suppressed~\cite{Anselmino:2009pn}, and the azimuthal phase (in the second term on the r.h.s.~of Eq.~(\ref{ds1})) oscillates strongly;
$ii)$ the Sivers effect does not suffer from any dynamical or azimuthal phase suppression. Indeed, in contrast to $pp\to\pi X$ processes in $\ell \, p \to \pi \, X$ only one partonic channel is at work and, for such moderate $Q^2$ values, the Sivers phase ($\phi$) in the first term on the r.h.s.~of Eq.~(\ref{ds1}) is still effective in the elementary interaction;
$iii)$ at this moderate c.m.~energy, even in the backward hemisphere of the polarized proton, one probes its valence region, where the extracted Sivers functions are sizeable and well constrained;
$iv)$ in the backward proton hemisphere at large $P_T$, $Q^2$ is predominantly larger than 1 GeV$^2$ and we can neglect any contribution from quasi-real photo-production.

\noindent
$\bullet$ Tagged or semi-inclusive category

We consider also the HERMES sub-sample data where the final lepton is tagged~\cite{Airapetian:2013bim} with $Q^2>1$ GeV$^2$, $W^2>10$ GeV$^2$, $0.023< x_B< 0.4$, $0.1<y<0.95$ and $0.2<z_h<0.7$ (standard SIDIS variables). We keep focusing only on the large $P_T$ region, namely $P_T > 1$ GeV.

\begin{figure}[h!t]
\includegraphics[width=3.9truecm,angle=0]{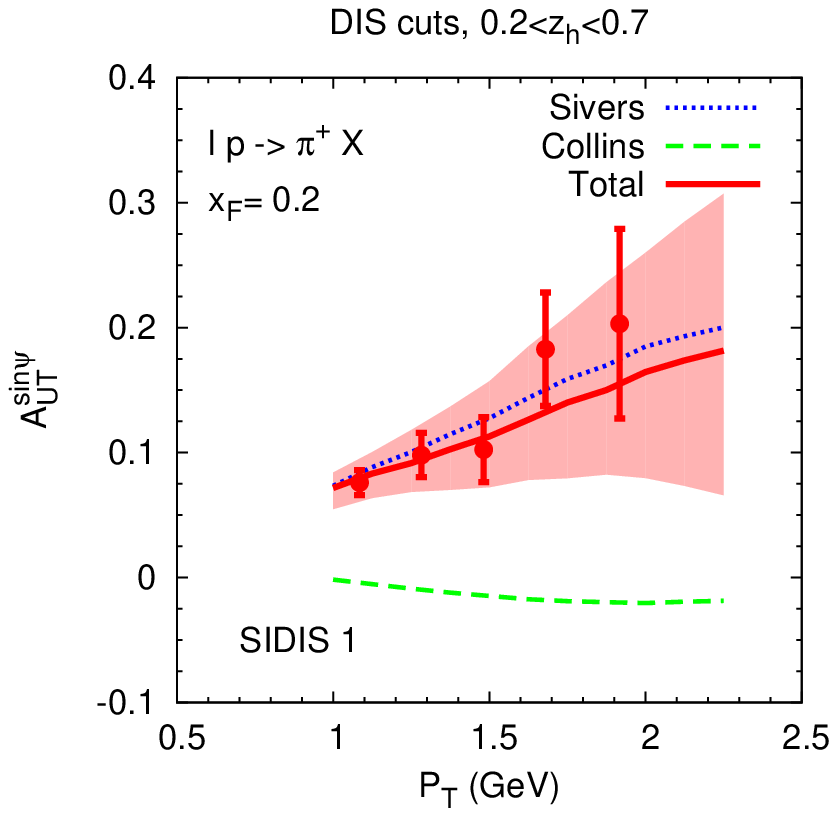}
\hspace*{-8pt}\includegraphics[width=3.9truecm,angle=0]{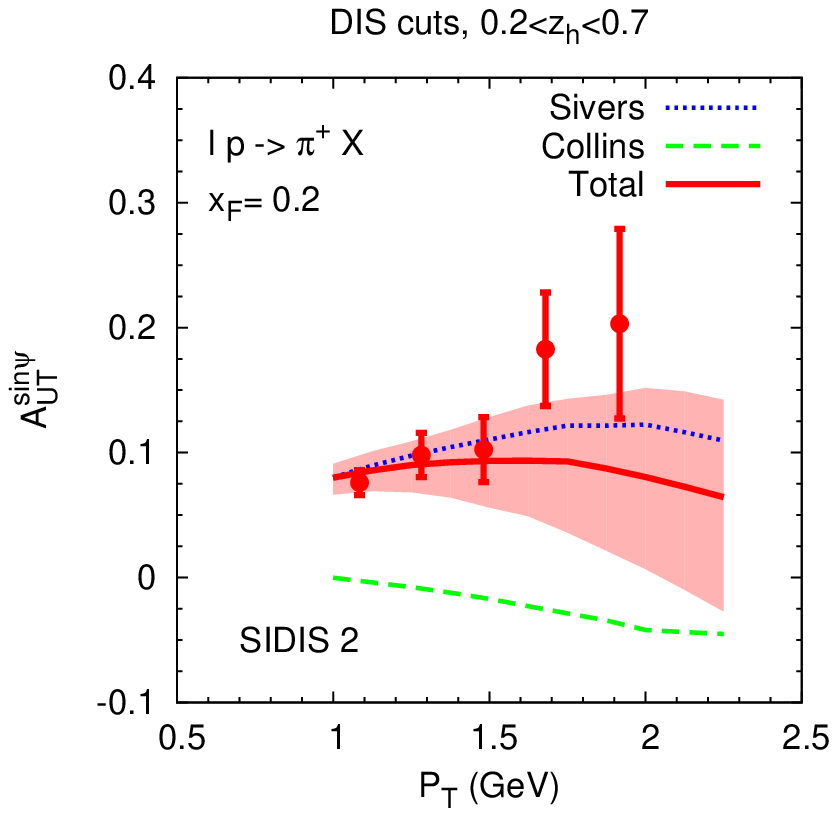}
\hspace*{-8pt}\includegraphics[width=3.9truecm,angle=0]{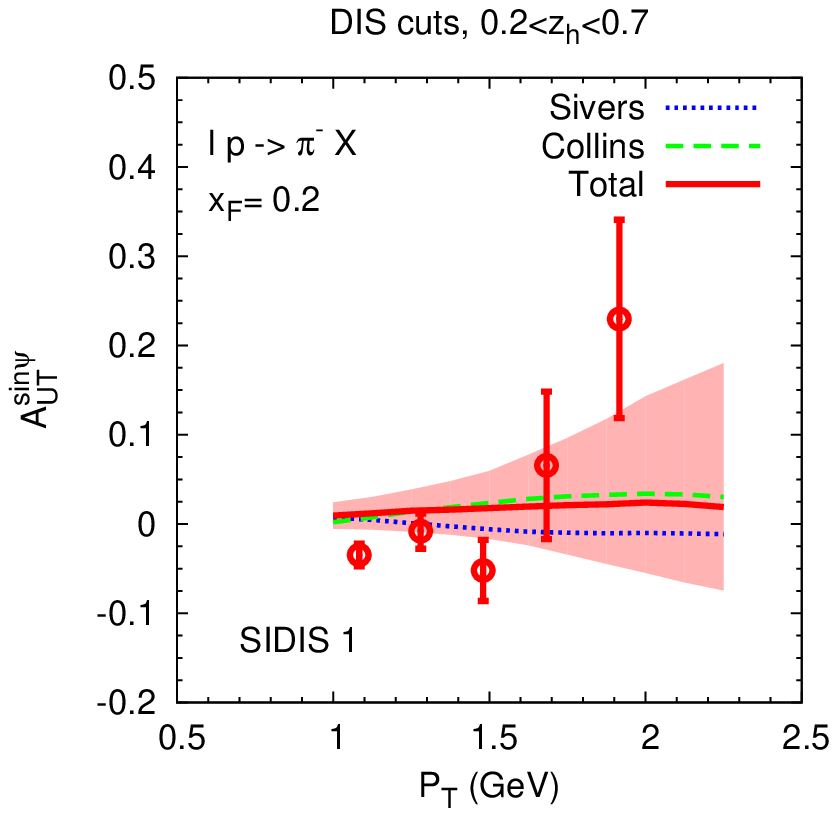}
\hspace*{-8pt}\includegraphics[width=3.9truecm,angle=0]{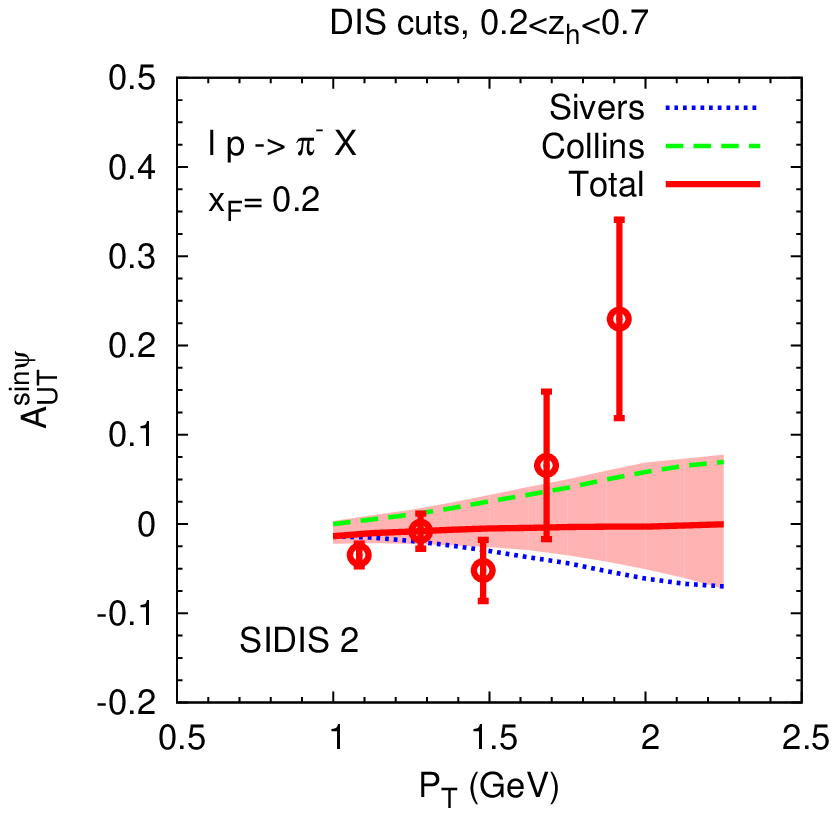}
\caption{
Theoretical estimates for $A_{UT}^{\sin\psi}$ vs.~$P_T$ at $x_F = 0.2$ for inclusive $\pi^+$ (first and second panel) and $\pi^-$ (third and fourth panel) production for the lepton tagged events in $\ell \, \pup \to \pi \, X$ process, computed according to Eqs.~(\protect\ref{AUT-hermes}) and (\protect\ref{anh})--(\protect\ref{ss1}) and compared
with the HERMES data~\cite{Airapetian:2013bim}.
\label{fig:an-hermes-dis-pi}}
\end{figure}

We show our estimates compared with HERMES data in Fig.~\ref{fig:an-hermes-dis-pi}, for positive and negative pion production as a function of $P_T$ at fixed $x_F = 0.2$. Again, we show the contributions from the Sivers (dotted blue line) and Collins
(dashed green line) effects separately and added together (solid red line) with the overall uncertainty bands (shaded area). Some comments follow:
$i)$ the Collins effect (dashed green lines) is only partially suppressed. The difference between the SIDIS 1 and the SIDIS 2 sets (a factor around 2-3) comes from the
different behaviour of the quark transversity functions at moderately large $x$;
$ii)$ the Sivers effect (dotted blue lines) for $\pi^+$ production (1st and 2nd panel) is sizeable for both sets. On the other hand for $\pi^-$ production the SIDIS 1 set (3rd panel) gives almost zero due to the strong cancellation between the unsuppressed Sivers up quark distribution coupled to the non-leading FF, with the more suppressed down quark distribution. For the SIDIS 2 set (4th panel), the same large $x$ behaviour of the up and down quark implies no cancellation.

The results expected for JLab 12 at $P_T \simeq 1$ GeV are similar to those observed at HERMES~\cite{Anselmino:2014eza}.
%
\begin{wrapfigure}{r}{0.51\columnwidth}
\vspace*{-.8cm}
\begin{center}
\hspace*{0.cm}
\includegraphics[width=3.9truecm,angle=0]{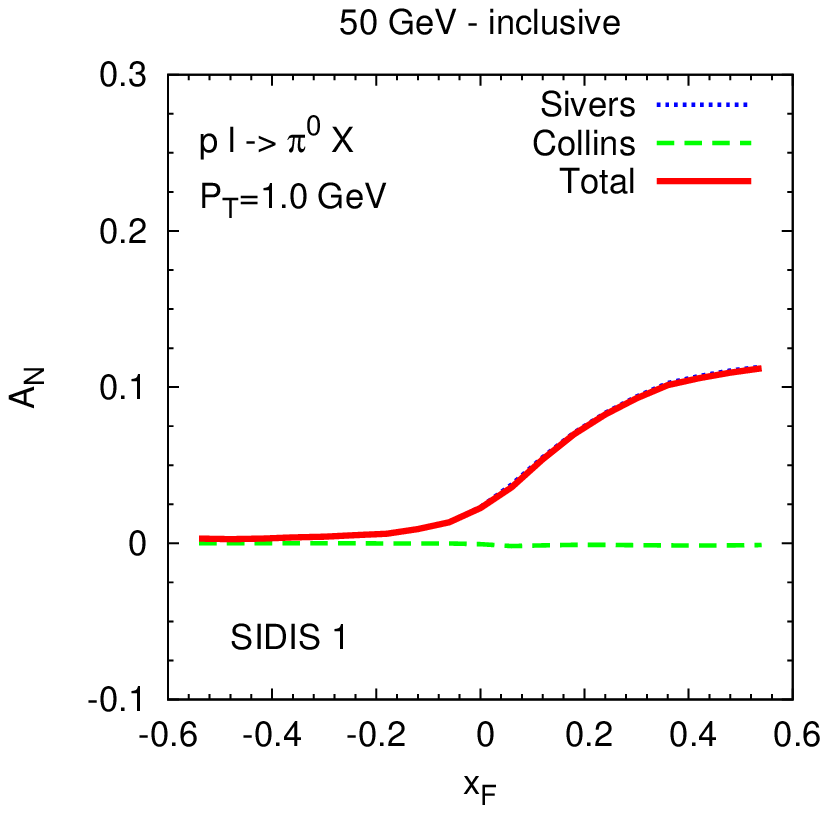}
\hspace*{-8pt}\includegraphics[width=3.9truecm,angle=0]{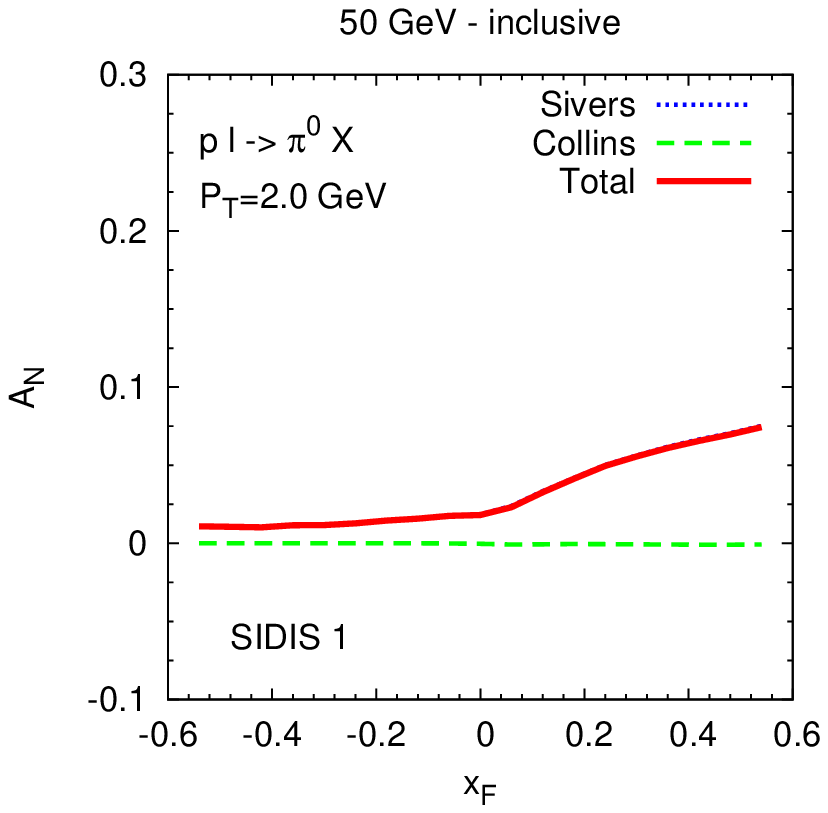}
\end{center}
\vspace*{-.5cm}
\caption{$A_{N}$ vs.~$x_F$ at $\sqrt{s}\simeq 50$ GeV,
$P_T=1$ GeV (left panel) and $P_T=2$ GeV (right panel) for $\pup \ell \to \pi^0 \, X$ (here a forward production w.r.t.~the proton direction corresponds to $x_F>0$).\label{eic}}
\end{wrapfigure}

\vspace*{-0.5cm}

Another interesting aspect is that at larger energies in a TMD scheme this process manifests some of the features of the SSAs in $p \, p \to \pi \, X$~\cite{Anselmino:2013rya, D'Alesio:2007jt}. Switching now to the configuration where the polarized proton moves along $Z_{cm}$, i.e.~with $x_F>0$ in the forward hemisphere of $\pup$, in Fig.~\ref{eic} we show some estimates of $A_N$ for $\pi^0$ production at $\sqrt s= 50$ GeV adopting the SIDIS 1 set (able to reproduces the behaviour of $A_N$ in $\pup p\to\pi\,X$ processes~\cite{Boglione:2007dm}). One can observe the following:
$i)$ the Collins effect in the backward region is totally negligible due to a strong suppression coming from the azimuthal phase integration. In the forward region both sets give tiny values; $ii)$ the Sivers effect is sizeable and increasing with $x_F$ for positive values of $x_F$, while negligible in the negative $x_F$ region. Even if there is only one partonic channel, the weak dependence on the azimuthal phase of the elementary interaction at the large $Q^2$ values reached at these energies implies a strong suppression at $x_F<0$.
Notice that this behaviour is similar to that observed at various energies in $A_N$ in $\pup p \to h \, X$ processes, being negligible at negative $x_F$ and increasing with positive values of $x_F$;
$iii)$ when one exploits the relation between the Qiu-Sterman function and the Sivers function the twist-3 approach for $A_N$ in $\ell \, \pup \to {\rm jet} + X$~\cite{Kang:2011jw} gives results similar, in sign and size, to those obtained in a TMD approach~\cite{Anselmino:2009pn}. However, the twist-3 collinear scheme, using the SIDIS Sivers functions, leads to values of $A_N$ in $pp\to \pi\, X$ collisions opposite to those measured~\cite{Kang:2011hk}.
A recent analysis of $A_N$ in $pp$ scattering in the twist-3 formalism~\cite{Kanazawa:2014dca} aiming at solving this problem introduces new large effects in the fragmentation. It is not clear how much these same effects would change the value of $A_N$ in $\ell p$ processes when going from jet to $\pi^0$ production;
$iv)$ the measurements of SSAs at such large energies, possible at a future Electron-Ion-Collider (EIC)~\cite{Accardi:2012qut} would be an invaluable tool to test the TMD factorization and discriminate among different approaches.

\end{document}